# First-principles calculations of structural, electronic and optical properties of HfZn$_2$

Md. Atikur Rahman[*1], Md. Afjalur Rahman[2], Md. Zahidur Rahaman[3]

[1, 2, 3] *Department of Physics, Pabna University of Science and Technology, Pabna-6600, Bangladesh*

## Abstract

The structural, electronic and optical properties of nearly ferromagnetic compound HfZn$_2$ have been studied using *ab-initio* technique. We have carried out the plane-wave pseudopotential approach within the framework of the first-principles density functional theory (DFT) implemented within the CASTEP code. The calculated structural parameters show a good agreement with the experimental results. In our work we have used GGA-PBE to calculate the electronic properties and it is found that the results exhibit similar band structure qualitatively with the results calculated using LDA. The electronic band structure reveals metallic conductivity and the major contribution comes from Hf-5d states. Our results for structural and electronic properties are compared with the experimental and other theoretical results wherever these are available. For the first time we have investigated the optical properties of HfZn$_2$ since no other experimental and theoretical studies on optical properties and dielectric functions of HfZn$_2$ has been reported yet. The reflectivity spectrum shows that the reflectivity is high in the visible-ultraviolet region up to 16 eV indicating promise as a good coating material to avoid solar heating.

## 1. Introduction

The Laves phase intermetallic compounds like HfZn$_2$ have attracted huge attention due to interesting characteristics including polymorphism and peculiar magnetic and electrical properties [1, 2]. The discovery of the ferromagnetic superconducting phase of ZrZn$_2$ has been attracted a great attention. The nearly ferromagnetic compound HfZn$_2$ has the same crystal structure with ZrZn$_2$. Due to this fact it should be quite interesting to study about HfZn$_2$ compound. According to experimental study HfZn$_2$ is an exchange enhanced paramagnet [3]. The susceptibility of HfZn$_2$ is about same as the susceptibility of the nearly ferromagnetic 4d element Pd though the material does not manage to reach the ferromagnetic state [4]. The linear specific heat coefficient for HfZn$_2$ is $\gamma = 15.8$ mJ/K$^2$ mole (formula unit) [4]. The large zero temperature magnetic susceptibility is much larger than that expected from the measured electronic specific heat coefficient $\gamma$, which indicates that exchange enhancement effects are sufficient enough to make the system ferromagnetic [13].

Computational methods using first principles have emerged a new era in condensed matter physics. Using this methods it is now possible to calculate the structural, electronic and optical properties of solids with good accuracy, which enabling us to explain and predict properties which are sometimes difficult to measure even in experimentally [5]. Now a day's first principles calculations based on density functional theory (DFT) have become an essential and important part of research in material science and condensed matter physics. Most of these theoretical studies are done by using pseudo-potential or FP- LAPW methods.

---

[*]Corresponding Author: atik0707phy@gmail.com

These theoretical studies are implemented using the DFT within the local density approximation (LDA) or generalized gradient approximation (GGA-PBE).To the best of our knowledge, theoretical investigations of HfZn$_2$ compound with GGA-PBE have been rarely reported yet.

In one literature published in 2006 [13], the electronic structure of HfZn$_2$ has been reported using LDA method. Therefore, a comprehensive study on structural and electronic properties of HfZn$_2$ using GGA-PBE method provides very useful information. Further, the optical properties of HfZn$_2$ have been reported yet to the best of our knowledge. The aim of this paper is to carry out a complete DFT-based search to investigate the structural, electronic and optical properties of nearly ferromagnetic compound HfZn$_2$. Calculated lattice parameters and band structure were methodically analyzed. In addition, a systematic comparison between the results obtained from this work and others works is also presented. Since there is no data available for the optical properties of HfZn$_2$, our study will help for experimental investigation on this compound in future. The remaining parts of this paper are organized as follows. In Section 2, the computation details are given. The results and discussion are presented in Section 3. Finally, a summary of our results is shown in Section 4.

## 2. Theoretical methods

The calculations have been carried out using the density functional theory (DFT) based CASTEP computer program together with the generalized gradient approximation (GGA) with the PBE exchange-correlation function [6-10]. Hf-5d$^2$6s$^2$ and Zn-3d$^{10}$4s$^2$ were taken as valence electrons. The electromagnetic wave functions are expanded in a plane wave basis set with an energy cut-off of 310 eV. The k-point sampling of the Brillouin zone was constructed using Monkhorst-Pack scheme [11] with 6×6×6 grids in primitive cells of HfZn$_2$. The equilibrium crystal structures were obtained via geometry optimization in the Broyden-Fletcher-Goldfarb-Shanno (BFGS) minimization scheme [12]. In the geometry optimization, criteria of convergence were set to 1.0×10$^{-5}$ eV/atom for energy, 0.03 eV/Å for force, 1×10$^{-3}$Å for ionic displacement, and 0.05 GPa for stress. These parameters are carefully tested and sufficient to lead to a well converged total energy.

## 3. Results and discussion

### 3.1. Structural properties

HfZn$_2$ intermetallic belongs to the family of C15 Laves phases of space group FD-3M (227). It has cubic crystal structure in which Hf atoms sit at 8a (0, 0, 0) and Zn atoms occupy Wyckoff position 16d (0.625, 0.625, 0.625).There are two formula units per unit cell. The lattice constants and atomic positions have been optimized as a function of normal stress by minimizing the total energy. The optimized structure is shown in Fig.1. The calculated values of the structural properties of HfZn$_2$ are presented in Table 1 along with the available experimental value and other theoretical results. From Table 1 we see that our present theoretical results are very close to both experimental result and other theoretical result. The calculated lattice constant of this present work is 7.375 Å which shows 0.74% deviation compared with the experimental value and slightly larger than other theoretical data due to the different calculation methods. This shows the reliability of our present DFT based first –principles calculations.

**Table 1.** Lattice constant "*a*", unit cell volume "*V*" and bulk modulus "*B*" for HfZn$_2$

| Properties | Expt.[13] | Others Calculation[14] | Present Calculation | Deviation from Expt. (%) |
|---|---|---|---|---|
| *a* (Å ) | 7.32 | 7.343 | 7.375 | 0.74 |
| V (Å$^3$/f.u.) | - | 49.52 | 50.141 | - |
| B (GPa) | - | 115.2 | 160.485 | - |

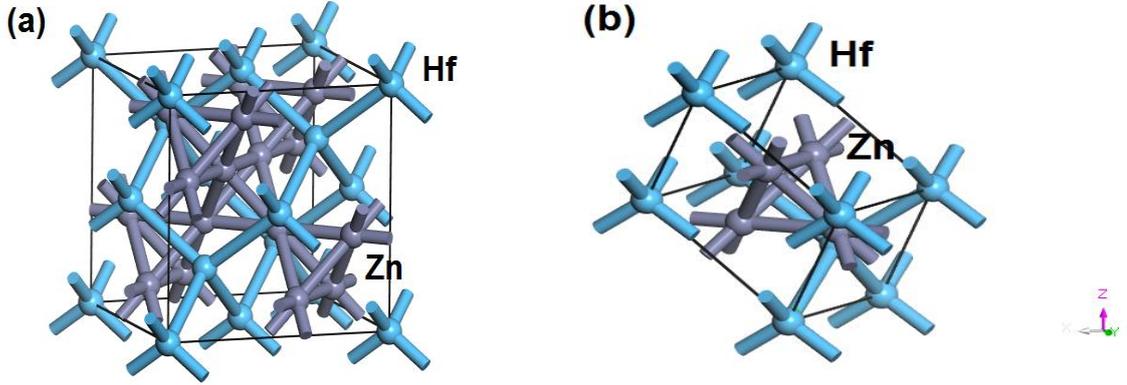

**Fig. 1.** The crystal structures of HfZn$_2$ (a) the conventional cubic cell and (b) the primitive cell.

### 3.2. Electronic properties

The density of states plays a vital role in the analysis of the physical properties of materials. Here we perform an analysis for the electronic band structure of HfZn$_2$. We illustrate the investigated partial density of states (PDOS) and total density of states (TDOS) at normal pressure. The calculated electronic band structures of HfZn$_2$ along the high symmetry directions in the Brillouin zones are shown in Fig. 2. The partial and total density of states of HfZn$_2$ is shown in Fig. 4. Here we have treated Hf-5d$^2$6s$^2$ and Zn-3d$^{10}$4s$^2$ as valence electrons.

It is observed from the band structure diagram shown in Fig. 3 that our investigated band structure is very much similar to that in Ref. [13] which is carried out using LDA method. This result confirms the reliability of the present calculations and also indicates that GGA-PBE method shows a good agreement with LDA method. The compound under study is metallic because a number of bands are overlapping at the Fermi level. Now to know about the details bonding characteristics we analyze PDOS diagram.

According to the PDOS and TDOS the lower valence band located at -8.0 eV to -6.0 eV is composed by Hf-5d, Hf-6s, Zn-3d and Zn-4s states, and it is dominated by Hf-6s and Zn-3d states. But Zn-3d state plays the main role to form this band. The middle valence band located at -6.0 eV to -5.0 eV originates from Hf-5d, Zn-3d and Zn-4s states where Hf-5d and Zn-4s states are dominant. The upper valence band located at -5.0 eV to 0 eV is made up of Hf-5d, Hf-6s and Zn-4s states where Hf-5d and Zn-4s states play the dominant role. In the conduction band, the Hf-5d state plays the dominant role from 0 eV to 1.5 eV. The region from 1.5 eV to 8 eV consists of Hf-5d, Hf-6s and Zn-4s states where Hf-6s and Zn-4s states are dominant. From the above analysis on the whole valence band, it is seen that Hf-6s states hybridize with the Zn-3d states in the energy range from -8.0 eV to -7.0 eV and Hf-5d states hybridize with the Zn-3d states in the energy range from -7.0 eV to -6.0 eV. In the energy range from -6.0 eV to -5.0 eV, Hf-5d states hybridize with Zn-4s states and from -5.0 eV to 0 eV Hf-5d states weakly hybridize with the Zn-4s states. The Hf-5d states play the dominant role near Fermi level ($E_F$). These hybridizations imply that the interatomic forces are central in HfZn$_2$ which is also confirmed in Ref.14.

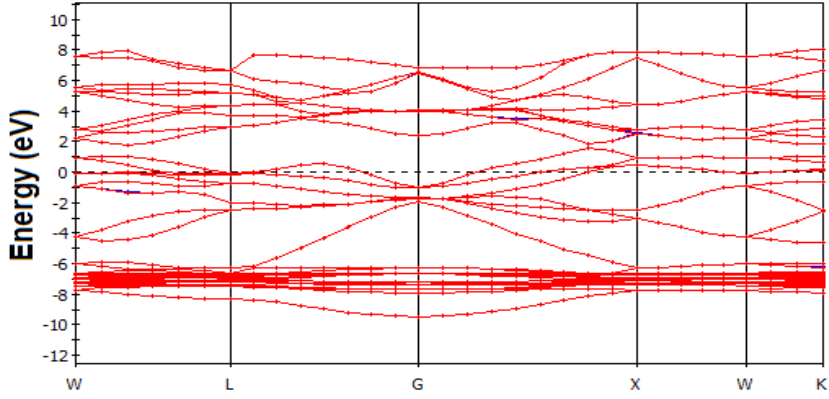

**Fig. 2.** Electronic band structure of HfZn$_2$ along the high symmetry directions in the Brillouin zones using GGA-PBE.

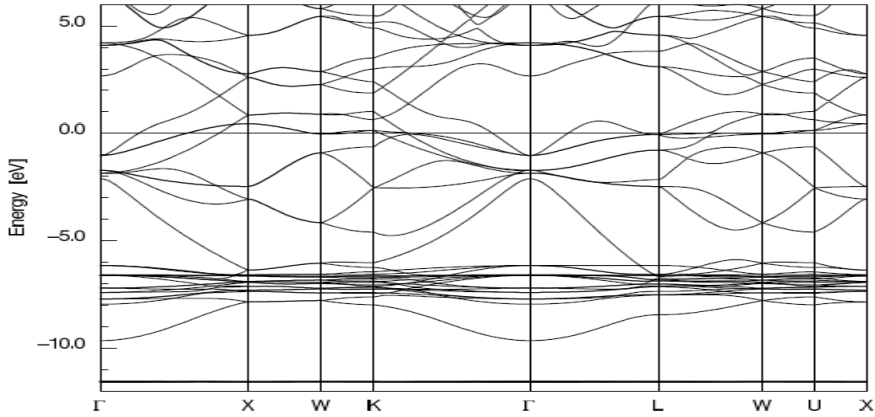

**Fig. 3.** Electronic band structure of HfZn$_2$ along the high symmetry directions in the Brillouin zones using LDA [13].

For further understanding about the bonding property, we have made further investigations on Mulliken overlap population [15] of HfZn$_2$. Mulliken overlap population is a great quantitative criterion for investigating the covalent and ionic nature of bonds. In Table 2, we have listed the atomic Mulliken population of HfZn$_2$. A high value of the bond population indicates a covalent bond, whereas a low value denotes ionic bonds. A value of zero indicates a perfectly ionic bond and the values greater than zero indicate the increasing levels of covalency [16]. From Table 2 we see that the bond populations are negative and certainly these are less than zero. Therefore, there exists no certain ionic or covalent bond rather these results indicates that the interatomic forces are central which has also confirmed in Ref. 14.

**Table 2.** Mulliken populations of HfZn₂ from GGA-PBE method.

| Species | s | p | d | Total | Charge | Bond | Population | Lengths |
|---------|------|------|------|-------|--------|---------|------------|----------|
| **Zn** | 0.31 | 1.77 | 9.93 | 12.01 | -0.01 | Zn1-Zn2 | -3.64 | 2.60772 |
| **Hf** | 0.63 | 0.61 | 2.73 | 3.97 | 0.03 | Zn1-Zn3 | -3.64 | 2.60772 |
| | | | | | | Zn2-Zn4 | -3.64 | 2.60772 |
| | | | | | | Zn2-Zn3 | -3.64 | 2.60772 |
| | | | | | | Zn1-Zn4 | -3.64 | 2.60772 |

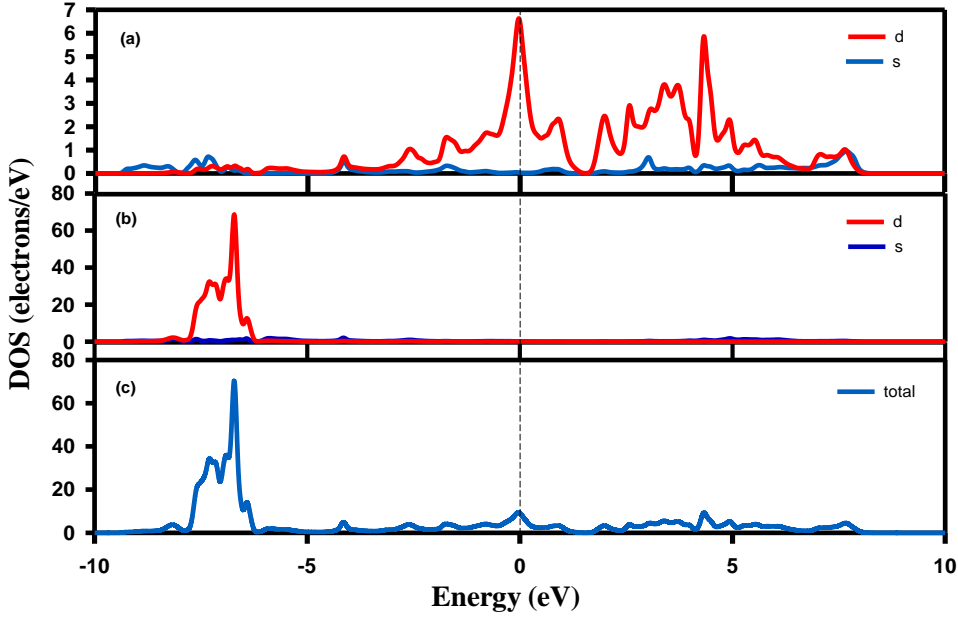

**Fig. 4.** Calculated partial and total density of states of HfZn₂ (a) for Hf (b) for Zn and (c) for total.

### 3.3. Optical properties

The study of the optical properties is crucial for understanding of the electronic structure of materials. These can be obtained from complex dielectric function ε (ω) which is defined as, *ε (ω) = ε₁ (ω) + iε₂ (ω)*. The imaginary part ε₂ (ω) is obtained from the momentum matrix elements between the occupied and the unoccupied electronic states and calculated directly by using Eq.1 [17],

$$\varepsilon_2(\omega) = \frac{2e^2\pi}{\Omega\varepsilon_0} \sum_{k,v,c} |\psi_k^c| u.r |\psi_k^v|^2 \delta(E_k^c - E_k^v - E) \qquad (1)$$

Where, *u* is the vector indicating the polarization of the incident electric field, *ω* is the frequency of light, *e* is the electronic charge and $\psi_k^c$ and $\psi_k^v$ are the conduction and valence band wave functions at *k*, respectively. Other optical constants, such as refractive index, loss-function, absorption spectrum, reflectivity and conductivity can be calculated by using Eq. 49 to 54 in ref. [17]. Fig.5 represents the optical functions of HfZn₂ calculated for photon energies up to 20 eV for polarization vector [100]. We have used a 0.5 eV Gaussain smearing for all calculations.

Fig. 5(a) shows the reflectivity spectra of HfZn$_2$ as a function of photon energy. We notice that the reflectivity is 0.40-0.75 in the infrared region and the value drops in the high energy region with some peaks. It is also found that the reflectivity of HfZn$_2$ is high in visible and ultraviolet region up to 16 eV (reaching maximum at 15-16 eV) which indicates the potentiality of this alloy to be used as a coating material to avoid solar heating.

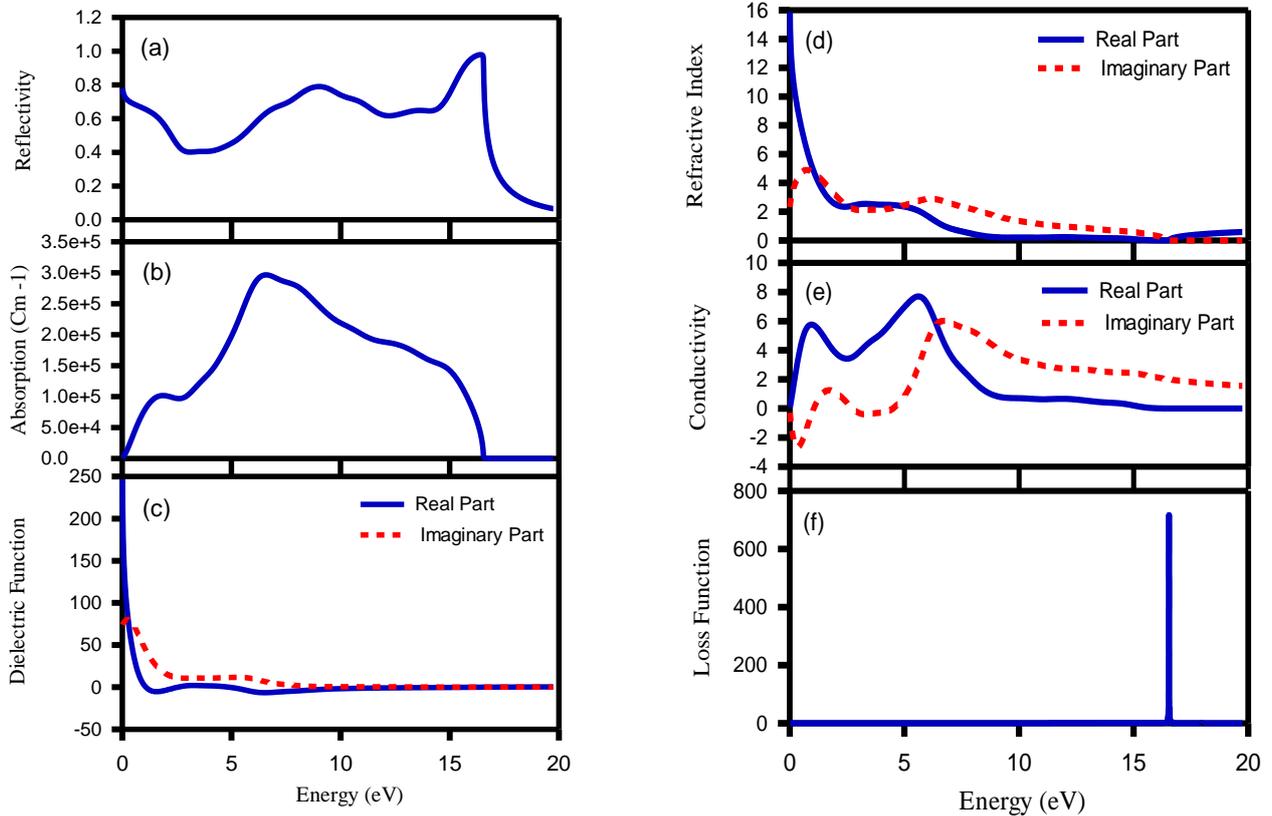

**Fig. 5.** The optical functions (a) reflectivity, (b) absorption, (c) dielectric function, (d) refractive index, (e) conductivity, and (f) loss function of HfZn$_2$ for polarization vector [100].

The absorption coefficient provides useful data about optimum solar energy conversion efficiency [18]. Fig. 5(b) exhibits the absorption coefficient spectra of HfZn$_2$ which reveal the metallic nature of this compound since the spectra starts from 0 eV. We observe two peaks, one at 1.80 eV and the other at 6.5 eV. This compound shows rather good absorption coefficient in the 5-10 eV region.

The dielectric function describes what an electric field such as an oscillating light wave does to material [19]. The quantity $\varepsilon_1 (\omega)$ represents how much a material becomes polarized when an electric field is applied due to creation of electric dipoles in the material while $\varepsilon_2 (\omega)$ represents absorption in a material [19]. When a material is transparent $\varepsilon_2 (\omega)$ is zero, but becomes nonzero when absorption begins. Fig 5(c) represents the dielectric function of HfZn$_2$ as a function of photon energy from which it is observed that the imaginary line approaches to zero at about 16.5 eV, indicating the transparency of this compound above 16.5 eV. Absorption occurs between 0-16.5 eV, which is also evident from Fig. 5(b). The value of the static dielectric constant of HfZn$_2$ is 250 which is quite large and can be used as a good dielectric material since materials with high dielectric constants are useful in the manufacture of high value capacitors [19]. We observe that the real part $\varepsilon_1 (\omega)$ of the dielectric function vanishes at about 16.5 eV. This corresponds to the energy at which the reflectivity exhibits a sharp drop and the energy loss function shows prominent peak. This peak in energy loss function at about 16.5 eV arises as $\varepsilon_1 (\omega)$ goes through zero and $\varepsilon_2 (\omega)$ is small at such energy, thus fulfilling the condition for plasma resonance at 16.5 eV. The material becomes transparent when the frequency of the incident light is higher than the plasma frequency.

The refractive index determines how much light is bent, or refracted, when entering a material [19]. The refractive index of $HfZn_2$ is represented in Fig. 5(d) as a function of photon energy from which we determine the static refractive index of $HfZn_2$ is 16. This is comparatively higher in the infrared region and gradually decreased in the visible and ultraviolet region.

Fig. 5(e) shows the conductivity spectra of $HfZn_2$ as a function of photon energy. We observed that the calculated optical conductivity have several maxima and minima within the energy range studied. Since the material has no band gap according to the band structure, the photoconductivity starts with zero photon energy as shown in Fig. 5(e). This result ensures the validity of our current DFT based calculations. Moreover, the photoconductivity and hence electrical conductivity of materials increases as a result of absorbing photons [20].

The electron energy loss function is an important optical parameter. Fig. 5(f) shows the energy loss function as a function of photon energy. Prominent peak is found at 16.5 eV, which indicates rapid reduction in the reflectance.

## 4. Conclusions

In summary, in this present work we have investigated the structural, electronic and optical properties of nearly ferromagnetic compound $HfZn_2$ by performing the generalized gradient approximation (GGA) in the frame of density functional theory. The calculated lattice parameters of the ground state structure show good agreement with the available experimental result. For the first time we have used GGA-PBE to calculate the band structures of $HfZn_2$ which shows good agreement with the band structures obtained from LDA method. We have also investigated the PDOS and TDOS of $HfZn_2$ compound and by analyzing those diagrams it is concluded that the interatomic forces are central in $HfZn_2$ which is also confirmed in Ref.14. The main aim of this work was to compare the results obtained from GGA-PBE and LDA method and we have done it successfully. From this study we indicate that GGA-PBE method shows a good agreement with LDA method.

For the first time we have studied the optical properties, e.g. absorption, conductivity, reflectivity, refractive index, energy-loss spectrum and dielectric function of $HfZn_2$. The reflectivity spectrum shows that the reflectivity is high in the visible-ultraviolet region up to 16 eV indicating promise as a good coating material. It is expected that our calculations should motivate experimental study on optical properties of $HfZn_2$ in future.

## References

[1] X.-Q. Chen, W. Wolf, R. Podloucky, P. Rogl, and M. Marsman, Phys. Rev. B 72, 054440 (2005).

[2] X. Zhang, L. Chen, M. Ma, Y. Zhu, S. Zhang, and R. Liu, J. Appl. Phys. 109, 113523 (2011).

[3] G.S.Knapp, B.W.Veal, and H.V.Cullbert, Inter. J. Magn. 1, 93 (1971).

[4] S.Foner and E.J.McNiff, Phys. Rev. Lett. 19, 1438 (1967).

[5]   R. Khenata, H. Baltache, M. Sahnoun, M. Driz, M. Rerat, B. Abbar, Physica B 336 (2003) 321.

[6]   S.J. Clark, M.D. Segall, C.J. Pickard, P.J. Hasnip, M.J. Probert, K. Refson, M.C. Payne, Z.Kristallogr. 220 (2005) 567–570.

[7]   Materials Studio CASTEP manual_Accelrys, 2010. pp. 261–262.
      <http://  www.tcm.phy.cam.ac.uk/castep/documentation/WebHelp/CASTEP.html>.

[8]   P. Hohenberg, W. Kohn, Phys. Rev. 136 (1964) B864–B871.

[9]   J.P. Perdew, A. Ruzsinszky, G.I. Csonka, O.A. Vydrov, G.E. Scuseria, L.A. Constantin, X. Zhou, K. Burke, Phys. Rev. Lett. 100 (2008) 136406–136409.

[10]  J.P. Perdew, A. Ruzsinszky, G.I. Csonka, O.A. Vydrov, G.E. Scuseria, L.A. Constantin, X. Zhou,  K. Burke, Phys. Rev. Lett. 100 (2008) 136406.

[11]  H. J. Monkhorst and J. D. Pack, Phys. Rev. B 13, 5188 (1976).

[12]  B. G. Pfrommer, M. Cote, S. G. Louie, and M. L. Cohen, J. Comput. Phys. 131, 233 (1997).

[13]  T. Jeong, Solid State Commun. 138, 265 (2006).

[14]  Sun, Na, et al. "First principle study of elastic and thermodynamic properties of $ZrZn_2$ and $HfZn_2$ under high pressure." *Journal of Applied Physics* 115.8 (2014): 083514.

[15]  R.S. Mulliken, J. Chem. Phys. 23 (1955) p.1833.

[16]  Segall, M. D.; Shah, R.; Pickard, C. J.; Payne, M. C. "Population analysis of plane-wave electronic structure calculations of bulk materials", *Phys. Rev. B*, 54, 16317-16320 (1996).

[17]  Materials Studio CASTEP manual_Accelrys, 2010.
      <http://www.tcm.phy.cam.ac.uk/castep/documentation/WebHelp/CASTEP.html>.

[18]  J. Sun, X.F. Zhou, Y.X. Fan, J. Chen, H.T. Wang, Phys. Rev. B 73 (2006) 045108 045110.

[19]  Rahman, Md, and Md Rahaman. "The structural, elastic, electronic and optical properties of MgCu under pressure: A first-principles study." *arXiv preprint arXiv:1510.02020* (2015).

[20]  Roknuzzaman, Md, and A. K. M. A. Islam. "Ab Initio Investigation of Nitride in Comparison with Carbide Phase of Superconducting InX (X= C, N)." *ISRN Condensed Matter Physics* 2013 (2013).